\documentclass[conference]{IEEEtran}
\IEEEoverridecommandlockouts
\usepackage{cite}
\usepackage{amsmath,amssymb,amsfonts}
\usepackage{graphicx}
\usepackage{textcomp}
\usepackage{xcolor}
\usepackage{siunitx}
\usepackage{booktabs}

\graphicspath{{Figures/}}

\begin{document}

\title{Toward Ku-Band Surface Acoustic Wave \\Delay Lines on AlScN-on-Diamond with \\Decoupled Phase and Group Velocities
\thanks{This work was supported by NSF CAREER 2339731: Radio Frequency Piezoelectric Acoustic Microsystems for Efficient and Adaptive Front-End Signal Processing.}}

\author{%
\IEEEauthorblockN{%
Tzu-Hsuan Hsu\textsuperscript{$*$}, Kapil Saha\textsuperscript{$\dagger$}, Yuchen Ma\textsuperscript{$*$}, Vakhtang Chulukhadze\textsuperscript{$*$}, Pietro Simeoni\textsuperscript{$\dagger$}, Matteo Rinaldi\textsuperscript{$\dagger$}, \\and Ruochen Lu\textsuperscript{$*$}}
\IEEEauthorblockA{%
Email: tzuhsuan.hsu@austin.utexas.edu\\ \textsuperscript{$*$}Department of Electrical and Computer Engineering, The University of Texas at Austin, Austin, TX, USA\\ \textsuperscript{$\dagger$}Department of Electrical and Computer Engineering, Northeastern University, Boston, MA, USA}}

\maketitle

\begin{abstract}
This work reports surface acoustic wave (SAW) acoustic delay lines (ADLs) on an aluminum scandium nitride (AlScN) on diamond platform operating in the X band and approaching the Ku band. The large acoustic-velocity contrast between the AlScN film and the diamond substrate produces a strongly dispersive Sezawa branch that decouples the phase velocity from the group velocity. Delay lines with a \SI{1}{\micro\meter} wavelength show a Sezawa passband at \SI{9.55}{\giga\hertz} with a fractional bandwidth of 1.05\%, a propagation loss of \SI{0.094}{\decibel} per wavelength, a propagation-limited quality factor of 465, and a group velocity of \SI{5979}{\meter\per\second}, while co-fabricated resonators give a phase velocity of \SI{9560}{\meter\per\second}, a ratio of about 1.6. Scaling the wavelength to \SI{0.5}{\micro\meter} moves the passband to \SI{16.5}{\giga\hertz} with a fractional bandwidth of 0.7\%. The platform therefore reaches an operating frequency about 1.5 times higher than AlScN on sapphire at the same lithographic pitch while preserving group delay per unit length.
\end{abstract}


\section{Introduction}

Wireless systems are moving toward bands above \SI{10}{\giga\hertz}, with the Ku band pivotal for next-generation systems \cite{andrews6g}, where acoustic components remain limited by material properties, mode confinement, and fabrication scalability \cite{gong2021}. Acoustic delay lines (ADLs) are attractive here because acoustic waves travel roughly five orders of magnitude slower than electromagnetic waves, so tens to hundreds of nanoseconds of delay fit on a die where an electromagnetic line yields about one nanosecond \cite{lu2019adl}. What the applications need, however, is not merely delay but a large amount of it. Self-interference cancellation in full-duplex radios must reproduce echoes arriving \SIrange{0.01}{1}{\micro\second} after transmission \cite{zhou2015}; radar and correlator front ends require delays matched to the chirp duration \cite{reindl2001}; and switched-delay-line nonreciprocal networks set their modulation frequency to the inverse of the delay, so a longer delay lowers the switching frequency and the power spent synthesizing it \cite{lu2019nonrecip, biedka2017}.

The useful figure is therefore delay per unit die length, and that figure is governed by the group velocity alone, since the delay of a line is its propagation length divided by $v_{g}$. Operating frequency, in contrast, is governed by the phase velocity, because the center frequency of an interdigital transducer (IDT) is $v_{p}/\lambda$. On a nondispersive platform, these are roughly the same number, and the two requirements create a trade-off. Raising the frequency through a faster substrate shortens the delay in proportion, while holding the delay fixed forces the wavelength down with IDT pitch bounded by patterning resolution and electrode loss \cite{hashimoto2000}.

\begin{figure}[!t]
\centerline{\includegraphics[width=0.88\columnwidth]{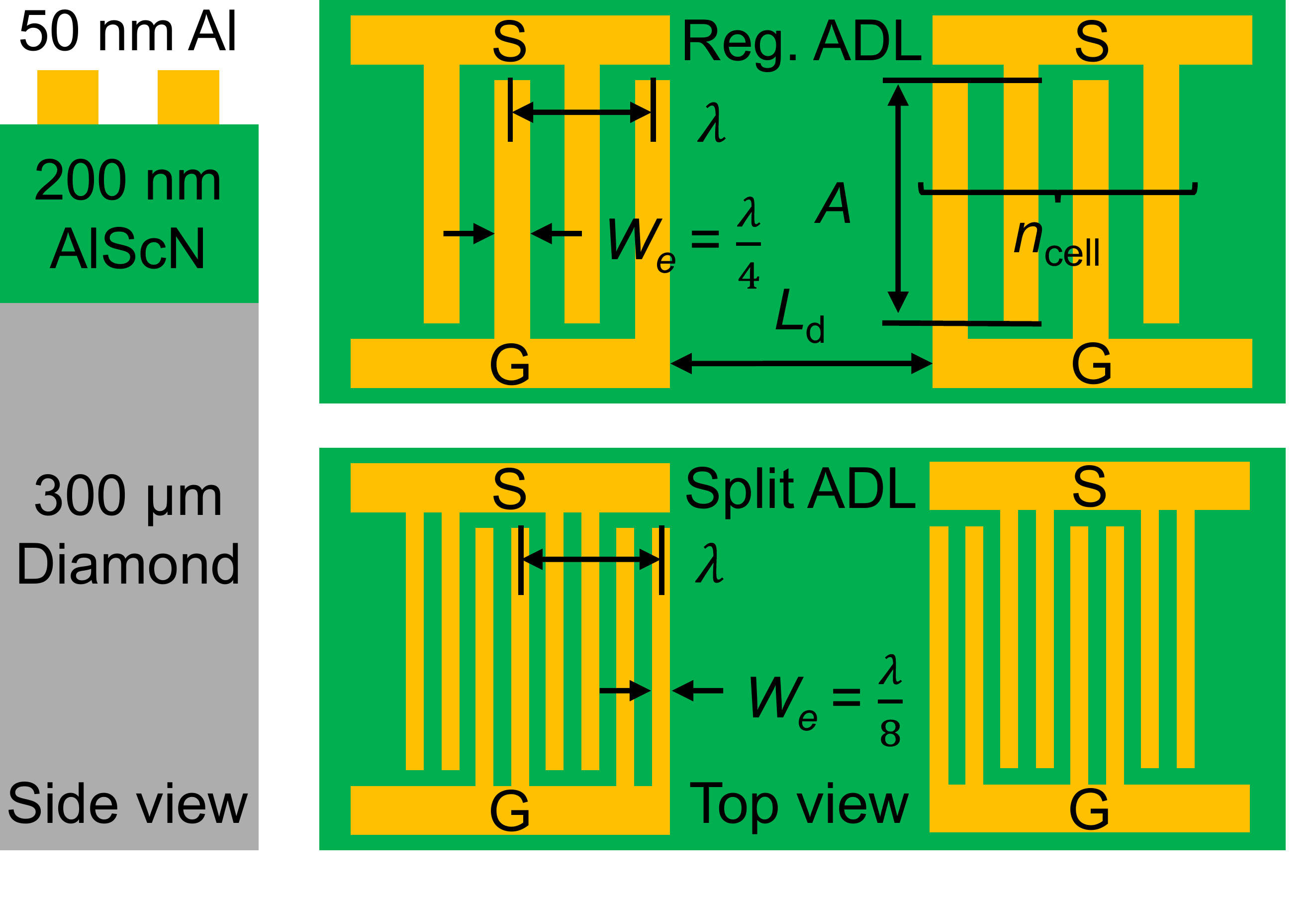}}
\caption{Side and top views of the AlScN-on-diamond SAW ADL, with the regular and split-electrode IDT topologies.}
\label{fig:stack}
\end{figure}

This work uses AlScN on diamond to break that velocity coupling. AlScN is retained for its excellent frequency scaling and fabrication flexibility \cite{cho2023}, while diamond offers the highest acoustic velocity of any bulk material and thermal conductivity above \SI{1500}{\watt\per\meter\per\kelvin}, with recent advancements in polycrystalline chemical vapor deposition (CVD) growth \cite{bolshakov2019} making it practical as a functional substrate \cite{hatashita2023}. The large velocity contrast confines the Sezawa mode and makes it strongly dispersive, so the phase velocity, set by the ratio of frequency to wavenumber, and the group velocity, set by the local slope of the dispersion relation, can be selected independently. We previously demonstrated the potential of the AlScN on diamond stack as a resonator platform \cite{hsu2025trans}. Here we further study the platform using acoustic delay lines, which access the group velocity and the propagation loss per wavelength that a resonance measurement leaves unresolved.

\begin{table}[!t]
\caption{Material Properties of AlScN and Candidate Substrates \cite{zhang2020, urban2021, zheng2021}}
\label{tab:materials}
\centering
\setlength{\tabcolsep}{2pt}
\renewcommand{\arraystretch}{1.1}
\footnotesize
\begin{tabular*}{\columnwidth}{@{\extracolsep{\fill}}lcccccc@{}}
\toprule
Material & $\rho$ & $C_{11}$ & $C_{44}$ & $v_{l}$ & $v_{s}$ & $\kappa$ \\
 & (kg/m$^{3}$) & (GPa) & (GPa) & (m/s) & (m/s) & (W/(m$\cdot$K)) \\
\midrule
Al$_{0.7}$Sc$_{0.3}$N & 3306 & 275  & 97   & 9120  & 5417  & 3.8  \\
Silicon               & 2329 & 166  & 79.6 & 8442  & 5846  & 142  \\
Sapphire              & 3968 & 490  & 145  & 11113 & 6045  & 32.5 \\
4H-SiC                & 3210 & 501  & 163  & 12493 & 7126  & 370  \\
\midrule
Diamond               & 3515 & 1079 & 578  & 17521 & 12823 & 1500 \\
\bottomrule
\end{tabular*}
\end{table}

\section{Design and Simulation}


Fig.~\ref{fig:stack} shows the device, a \SI{200}{\nano\meter} sputtered Al$_{0.7}$Sc$_{0.3}$N film on a \SI{300}{\micro\meter} polycrystalline CVD diamond substrate with \SI{50}{\nano\meter} aluminum IDTs. As depicted in Table~\ref{tab:materials}, diamond exceeds sapphire and 4H-SiC in both longitudinal and shear velocities, making it an ideal material to form an acoustic waveguide. Aluminum electrodes are selected to add minimal mass loading to the Sezawa mode. Unlike platforms that rely on dense electrodes to build velocity contrast \cite{du2024, hao2019}, diamond confines wave propagation directly. Two IDT topologies are studied. The regular ADL uses $W_{e} = \lambda/4$, and the split-electrode ADL uses $W_{e} = \lambda/8$, which suppresses internal reflection at the cost of finer features. Both use an aperture of $40\lambda$ and 30 cells, with delay lengths $L_{d}$ of 80, 160, 240, and 320 wavelengths fabricated on the same die to extract loss and group velocity from a regression fit. 


\begin{figure}[!t]
\centerline{\includegraphics[width=0.88\columnwidth]{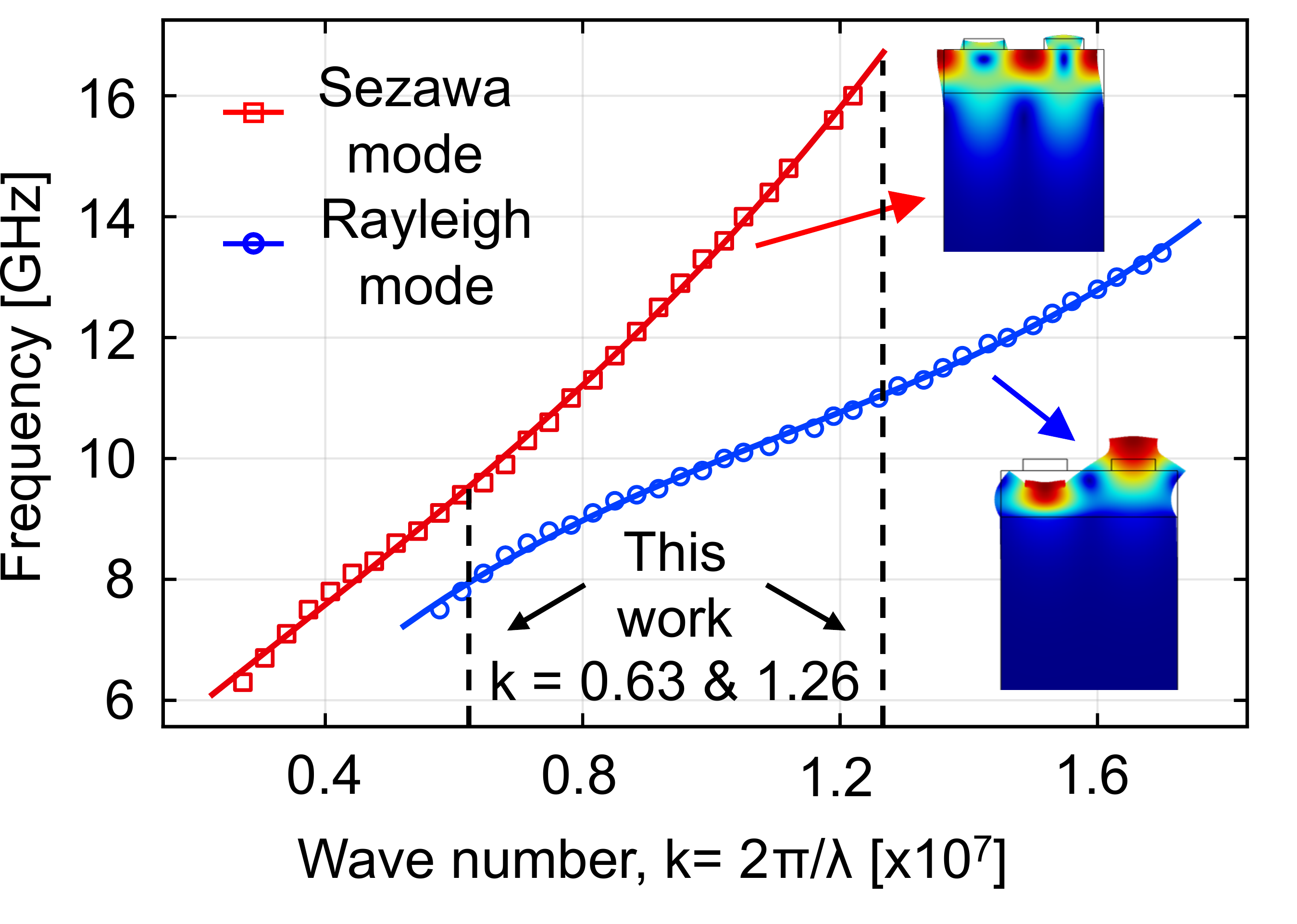}}
\caption{Simulated dispersion of Rayleigh and Sezawa modes, with inset showing mode shapes. The marked region spans $\lambda = \SI{1}{\micro\meter}$ to $\SI{0.5}{\micro\meter}$.}
\label{fig:dispersion}
\end{figure}

The velocity dispersion was first simulated in COMSOL using a unit cell model with periodic boundaries \cite{hsu2021jmm} terminated by a perfectly matched layer at the bottom with loss factors omitted. Fig.~\ref{fig:dispersion} plots the first two guided branches against wavenumber $k = 2\pi/\lambda$. The phase velocity is the secant slope from the origin, $v_{p} = \omega/k$, and the group velocity is the local tangent slope, $v_{g} = \partial \omega / \partial k$. A nondispersive mode implies a straight line through the origin so that the two velocities coincide. The Sezawa branch is displaced upward and curved, so its secant slope exceeds its tangent slope over the marked region, which spans the two design points at $\lambda = \SI{1}{\micro\meter}$ and $\lambda = \SI{0.5}{\micro\meter}$. Evaluating both slopes at $\lambda = \SI{1}{\micro\meter}$ predicts $v_{p}$ of \SI{9578}{\meter\per\second} and $v_{g}$ of \SI{5727}{\meter\per\second}, a ratio of 1.67.


\begin{figure}[!t]
\centerline{\includegraphics[width=0.88\columnwidth]{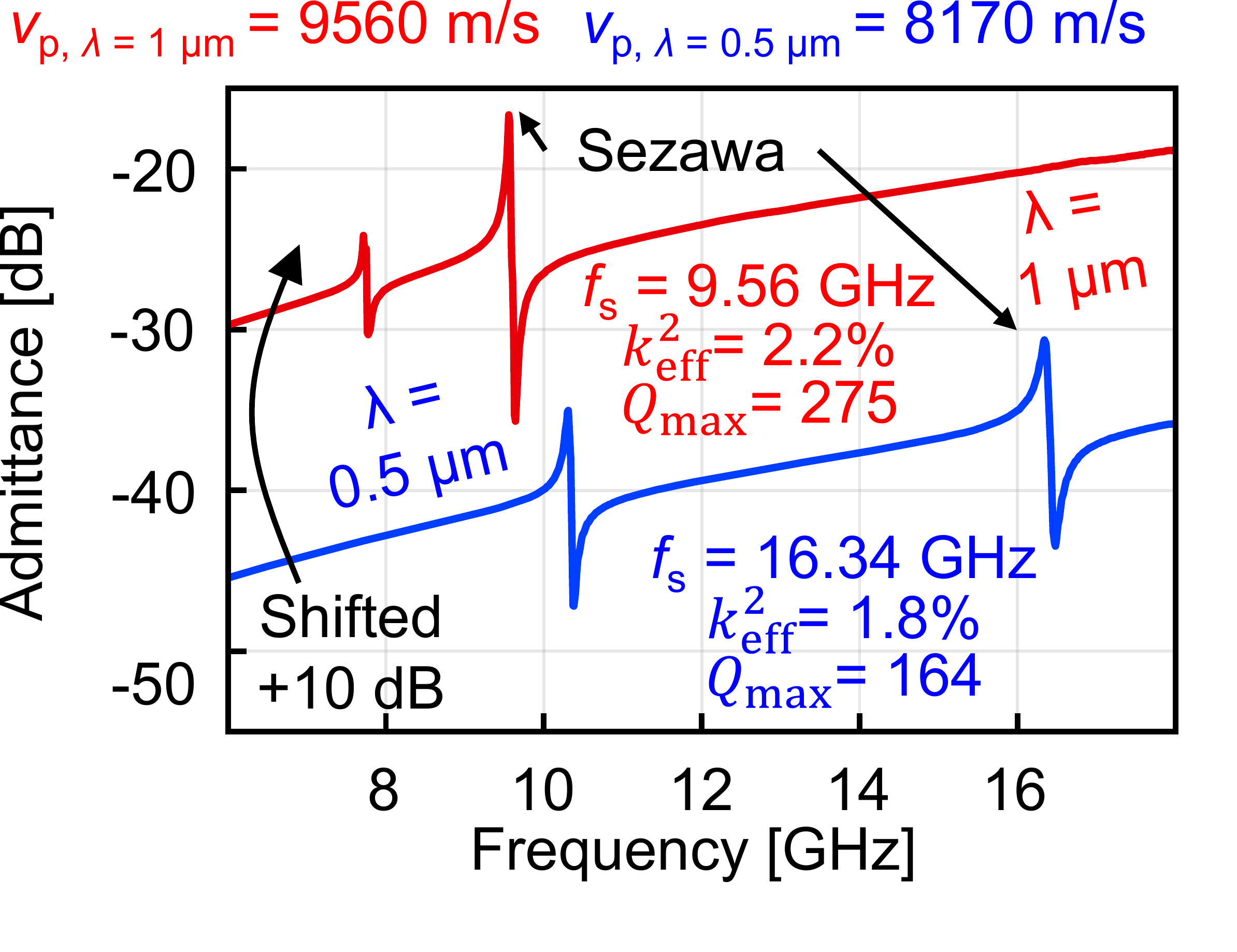}}
\caption{Measured admittance of co-fabricated resonators anchoring $v_{p}$ at both design points. Traces shifted for clarity.}
\label{fig:resvp}
\end{figure}

\begin{figure}[!t]
\centerline{\includegraphics[width=0.8\columnwidth]{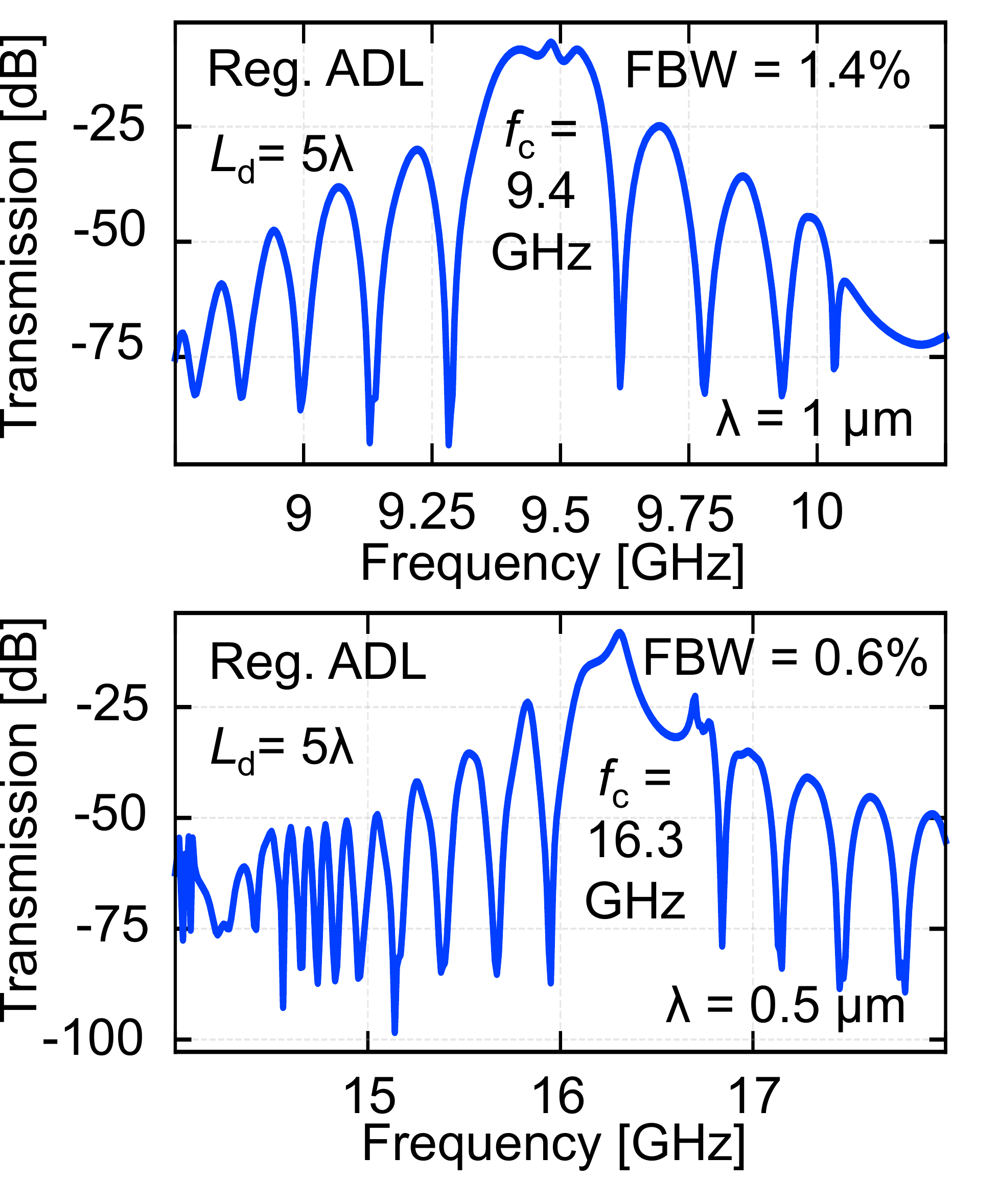}}
\caption{Simulated transmission of the regular ADL with a five-wavelength delay at $\lambda = \SI{1}{\micro\meter}$ (top) and \SI{0.5}{\micro\meter} (bottom).}
\label{fig:simulation}
\end{figure}

Resonators co-fabricated on the same die confirm the simulated velocity results. Fig.~\ref{fig:resvp} shows the measured admittance, taken with the setup of Section IV. The coupling is extracted from the series and parallel resonances, and the Bode quality factor from the measured reflection coefficient \cite{hsu2020edl, jin2021},
\begin{equation}
k_{\mathrm{eff}}^{2} = \frac{\pi^{2}}{8}\frac{f_{p}^{2}-f_{s}^{2}}{f_{s}^{2}}, \qquad
Q_{\mathrm{Bode}} = \omega \left| \frac{dS_{11}}{d\omega} \right| \frac{1}{1-|S_{11}|^{2}}.
\label{eq:k2q}
\end{equation}
At $\lambda = \SI{1}{\micro\meter}$ the Sezawa mode resonates at \SI{9.56}{\giga\hertz} with $k_{\mathrm{eff}}^{2}$ of 2.2\% and $Q_{\max}$ of 275, giving $v_{p} = \SI{9560}{\meter\per\second}$; at $\lambda = \SI{0.5}{\micro\meter}$ it resonates at \SI{16.34}{\giga\hertz} with $k_{\mathrm{eff}}^{2}$ of 1.8\% and $Q_{\max}$ of 164, giving \SI{8170}{\meter\per\second}. This fall with halved wavelength is the direct signature of the curvature in Fig.~\ref{fig:dispersion}, and it brackets the \SI{8671}{\meter\per\second} measured previously at an intermediate thickness on this stack \cite{hsu2025trans}, placing three measurements from two fabrication runs on one branch.


Fig.~\ref{fig:simulation} shows the simulated transmission of the regular ADL. A five-wavelength delay is used so the response isolates transduction, and these models are not intended to predict the insertion loss of the fabricated devices, whose delays are more than an order of magnitude longer. The simulated centers of \SI{9.4}{\giga\hertz} and \SI{16.3}{\giga\hertz} match the measured passbands of Section IV to within 2\%.

\section{Fabrication}

Devices were fabricated on a \SI{300}{\micro\meter} polycrystalline CVD diamond substrate. Diamond is difficult to machine, and substrate roughness degrades film crystallinity; the substrate here has $R_{a}$ below \SI{20}{\nano\meter}. A \SI{200}{\nano\meter} Al$_{0.7}$Sc$_{0.3}$N film was sputtered onto the diamond, with an X-ray diffraction rocking curve full width at half maximum of \ang{3.42} \cite{hsu2025trans}. \SI{50}{\nano\meter} aluminum electrodes were then patterned using electron beam lithography followed by lift-off to define the IDTs and pads.

\section{Measurement Results}

\begin{figure}[!t]
\centerline{\includegraphics[width=0.80\columnwidth]{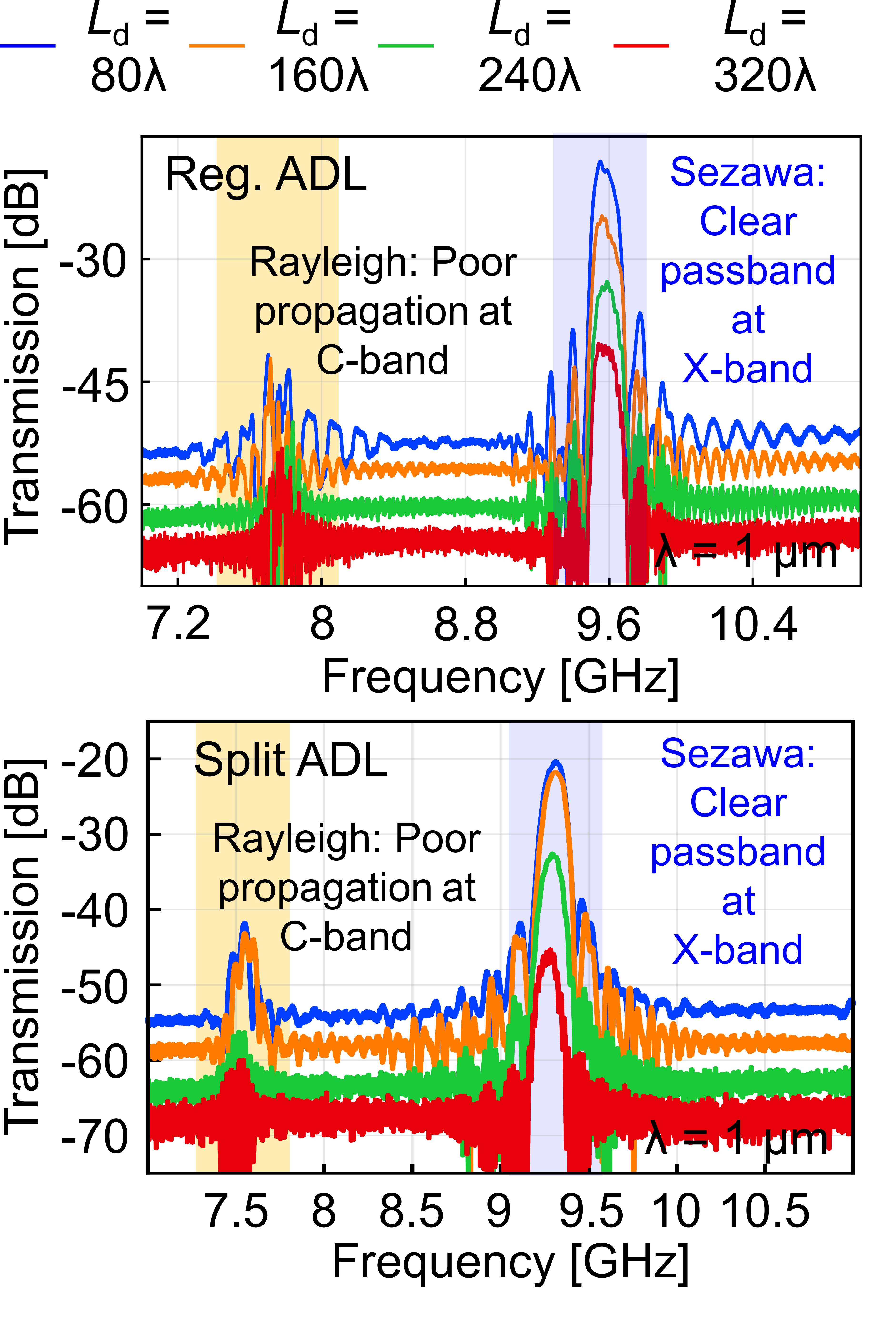}}
\caption{Measured wideband transmission at $\lambda = \SI{1}{\micro\meter}$ for $L_{d}$ of 80 to 320 wavelengths, regular (top) and split-electrode (bottom) ADLs.}
\label{fig:wideband}
\end{figure}


Devices were measured on a Keysight P5028A vector network analyzer with an MPI TS150 probe station under ambient conditions, after a standard short-open-load-through calibration. All data are shown as measured, without de-embedding. Fig.~\ref{fig:wideband} shows the wideband transmission of both topologies at $\lambda = \SI{1}{\micro\meter}$. The Sezawa mode forms a clear passband near \SI{9.5}{\giga\hertz} whose peak transmission falls monotonically with delay length. The Rayleigh mode near \SI{7.5}{\giga\hertz} shows only weak transmission that does not scale cleanly, indicating poor propagation at C-band. This asymmetry is invisible to a resonator, where both modes resonate cleanly \cite{hsu2025trans}.


\begin{figure}[!t]
\centerline{\includegraphics[width=0.9\columnwidth]{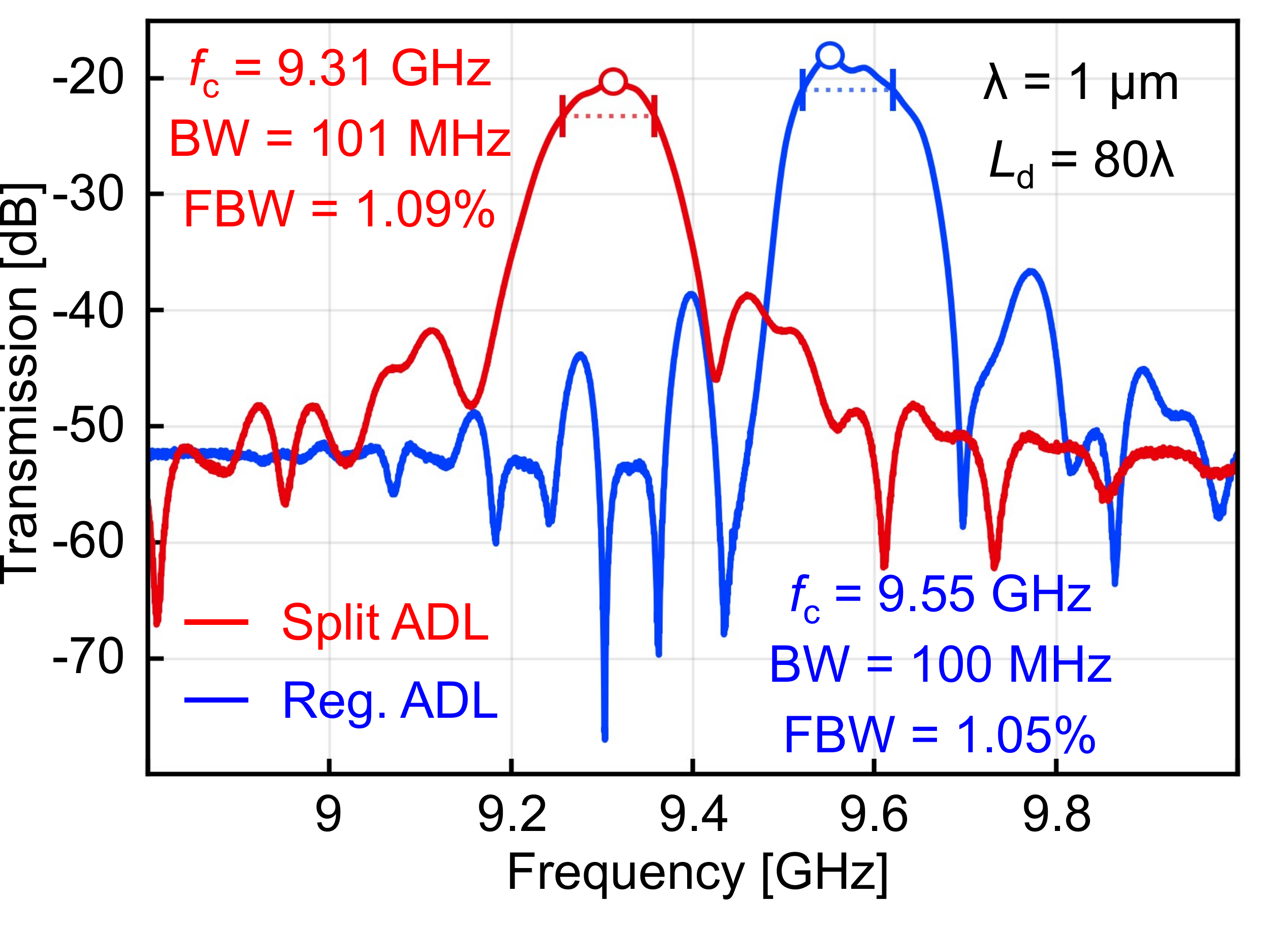}}
\caption{Measured Sezawa passband at $\lambda = \SI{1}{\micro\meter}$ and $L_{d} = 80\lambda$ for the regular and split-electrode ADLs.}
\label{fig:passband}
\end{figure}

Fig.~\ref{fig:passband} compares the topologies at $L_{d} = 80\lambda$. The regular ADL is centered at \SI{9.55}{\giga\hertz} with a \SI{3}{\decibel} bandwidth of \SI{100}{\mega\hertz}, a fractional bandwidth of 1.05\%, and a minimum insertion loss of \SI{18}{\decibel}; the split-electrode ADL gives \SI{9.31}{\giga\hertz}, \SI{101}{\mega\hertz}, 1.09\%, and \SI{20}{\decibel}. The two topologies show essentially identical bandwidth, with the split design sitting \SI{240}{\mega\hertz} lower and showing a cleaner passband and sidelobes.


\begin{figure}[!t]
\centerline{\includegraphics[width=0.88\columnwidth]{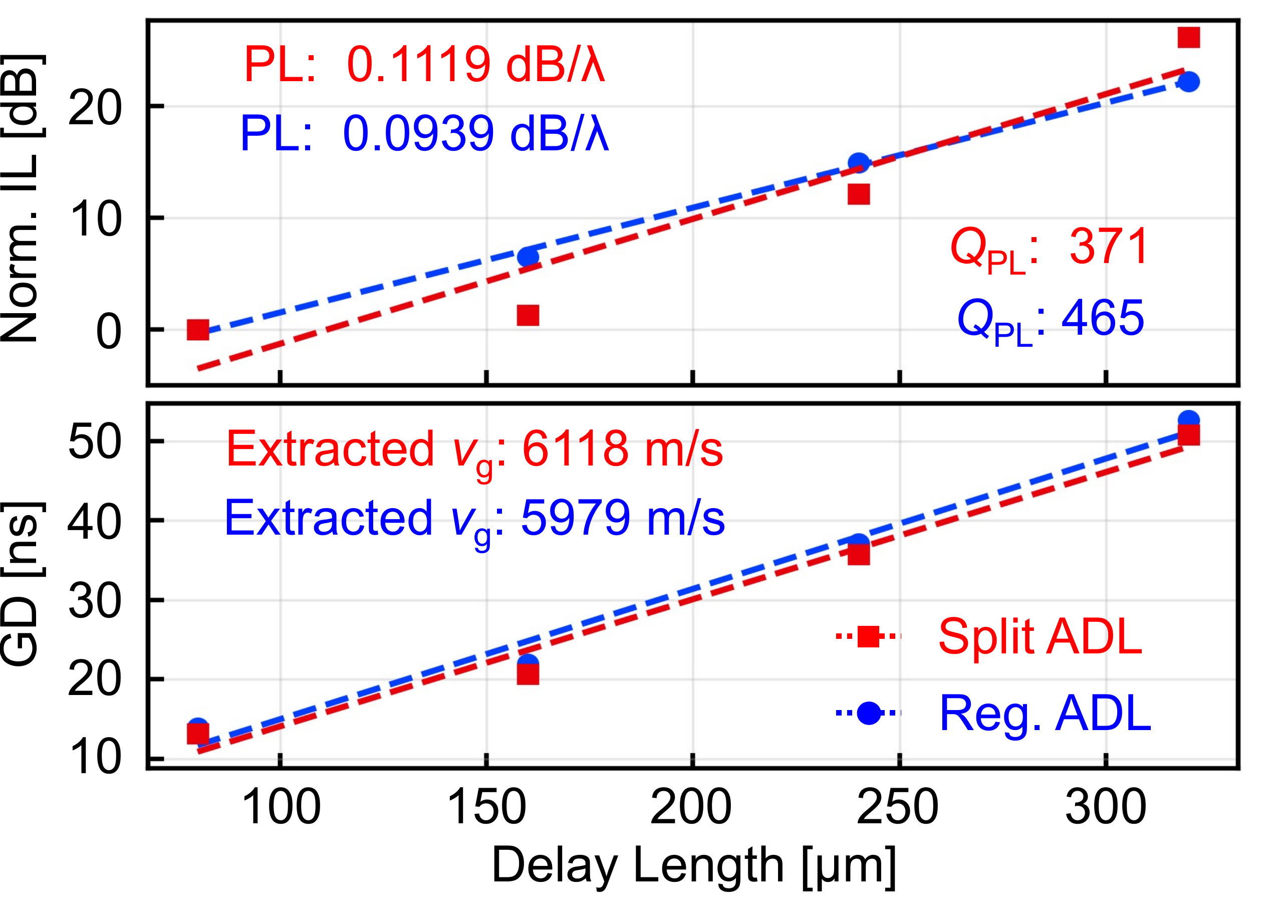}}
\caption{Normalized insertion loss (top) and group delay (bottom) against delay length, giving propagation loss, $Q_{\mathrm{PL}}$, and $v_{g}$ by regression.}
\label{fig:plext}
\end{figure}

Fig.~\ref{fig:plext} shows the propagation loss extraction across all delay lengths, following the method of \cite{lu2019adl}. As expected, the normalized insertion loss rises linearly, giving \SI{0.0939}{\decibel} per wavelength for the regular ADL and \SI{0.1119}{\decibel} for the split-electrode ADL, with $Q_{\mathrm{PL}}$ of 465 and 371, respectively. The regular topology has the lower loss of the two, consistent with its coarser features. Group delay also rises linearly, from about \SI{13}{\nano\second} at $80\lambda$ to \SI{52}{\nano\second} at $320\lambda$, giving group velocities of \SI{5979}{\meter\per\second} and \SI{6118}{\meter\per\second}, both close to the simulated $v_{g}$.

By using the extracted $v_{p} = \SI{9560}{\meter\per\second}$, the regular ADL gives a velocity ratio of $v_{p}/v_{g} = 1.60$. By contrast, a typical nondispersive platform reaching \SI{9.55}{\giga\hertz} at the same wavelength would propagate at \SI{9560}{\meter\per\second} and deliver 37\% less group delay over the same die length.

Currently, the $Q_{\mathrm{PL}}$ of 465 reflects process maturity rather than a fundamental limit. The Sezawa mode penetrates deep into the substrate \cite{hsu2025trans}, making it sensitive to scattering from the diamond grain structure and the AlScN interface, whose roughness is a tenth of the film thickness here.


\begin{figure}[!t]
\centerline{\includegraphics[width=0.88\columnwidth]{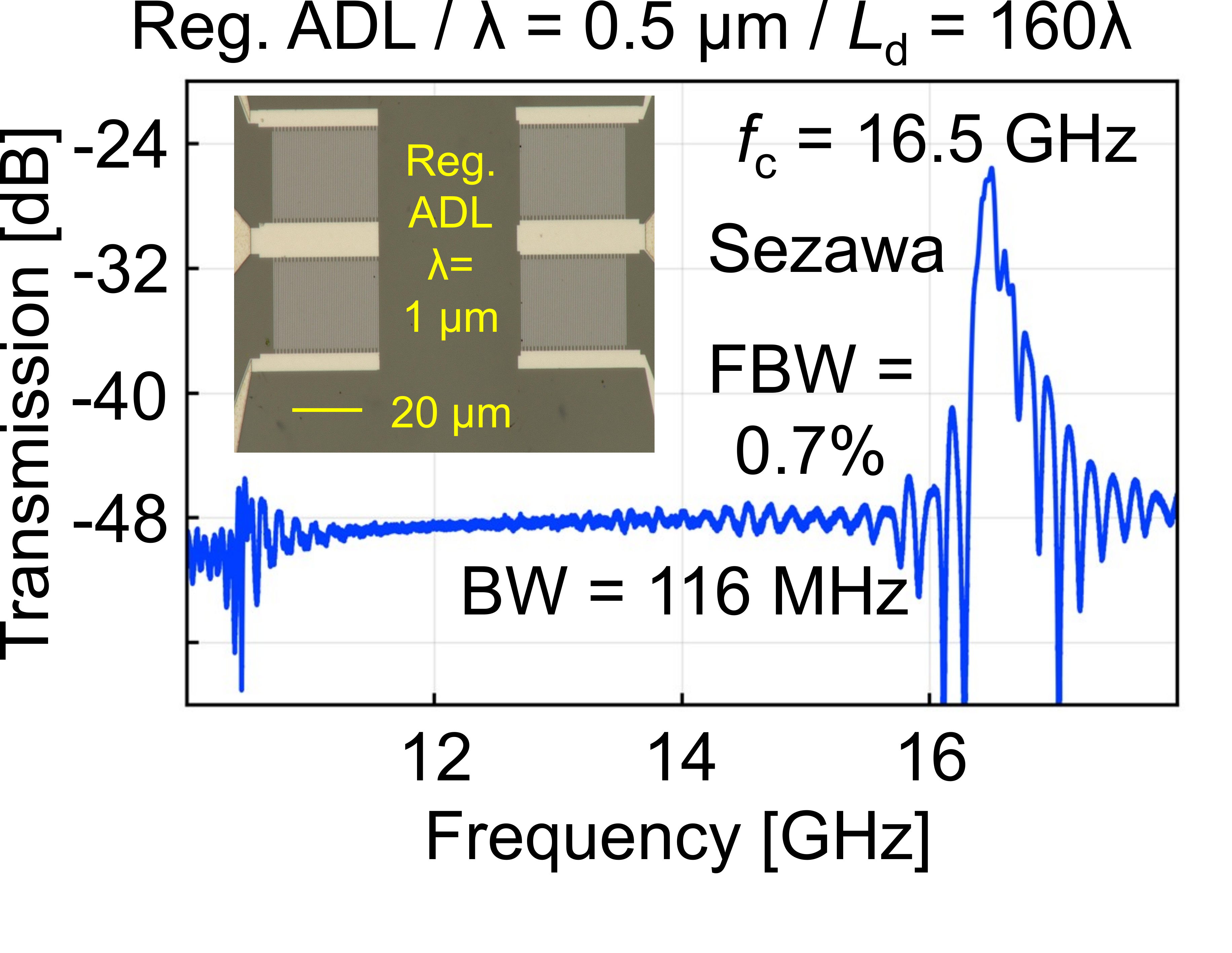}}
\caption{Measured Sezawa passband of the $\lambda = \SI{0.5}{\micro\meter}$ ADL at $L_{d} = 160\lambda$.}
\label{fig:hfadl}
\end{figure}

To further explore higher frequency scalability, Fig.~\ref{fig:hfadl} shows a device with $\lambda = \SI{0.5}{\micro\meter}$ and $L_{d} = 160\lambda$, with a Sezawa passband at \SI{16.5}{\giga\hertz}, a \SI{3}{\decibel} bandwidth of \SI{116}{\mega\hertz}, and a fractional bandwidth of 0.7\%, inside the Ku band. Halving the wavelength raises the frequency by a factor of 1.73. At the same pitch, an AlScN-on-sapphire platform would sit about 1.5 times lower in frequency, and reaching a comparable frequency on SiC required up to a 20\% smaller feature size \cite{colombo2025}. For delay lines, this relaxed patterning comes without the loss of delay density that a faster nondispersive platform would incur.

\section{Conclusion}

In this work, highly velocity-dispersive SAW delay lines have been demonstrated on AlScN-on-diamond at \SI{9.55}{\giga\hertz} and \SI{16.5}{\giga\hertz}. The dispersive Sezawa branch decouples phase velocity from group velocity, and the measured ratio of 1.60 shows that frequency can be raised without a proportional loss of delay per unit length. Propagation loss of \SI{0.094}{\decibel} per wavelength and $Q_{\mathrm{PL}}$ of 465 reflect sub-optimal diamond surface conditions prior to AlScN growth and require further optimization. Overall, the results establish this platform for frequency-scalable delay elements in the X and Ku bands.



\end{document}